\documentclass[pdf]{iucr}
\usepackage{amsmath}
\usepackage{amssymb}
\usepackage{amsthm}
\usepackage{xcolor}
\usepackage{bm}
\usepackage[normalem]{ulem}
\usepackage{hyperref}
\usepackage{lineno}

\newcommand{\un}[1]{\ensuremath{\,\mathrm{#1}}}
\newcommand{\chem}[1]{\ensuremath{\mathrm{#1}}}

\newcommand{\tensor}[1]{\ensuremath{\boldsymbol{#1}}}
\renewcommand{\vec}[1]{\ensuremath{\boldsymbol{#1}}}

\newcommand{\lab}[0]{\ensuremath{\ell}}
\newcommand{\diff}[0]{\ensuremath{s}}

\newcommand{\sym}[0]{\ensuremath{\mathrm{sym}}}

\journalcode{J} 

\begin{document}


\title{Oblique diffraction geometry for the observation of several non-coplanar Bragg reflections under identical illumination}

\cauthor[a]{C.}{Detlefs}{detlefs@esrf.fr}{}
\author[b]{A.}{Henningsson}
\author[c,d]{B.}{Kanesalingam}
\author[e]{A. A. W.}{Cretton}
\author[a]{C.}{Corley-Wiciak}
\author[b]{F. T.}{Frankus}
\author[c,d]{D.}{Pal}
\author[c,d]{S.}{Irvine}
\author[e]{S.}{Borgi}
\author[e]{H. F.}{Poulsen}
\author[a]{C.}{Yildirim}
\author[c,d]{L. E.}{Dresselhaus-Marais}
 
\aff[a]{European Synchrotron Radiation Facility, 38043 \city{Grenoble}, \country{France}}

\aff[b]{Technical University of Denmark, Department of Mechanical Engineering, \city{Kgs. Lyngby}, \country{Denmark}}
 
\aff[c]{Stanford University, Department of Materials Science and Engineering, California 94305, \country{USA}}

\aff[d]{SLAC National Accelerator Laboratory, California 94025, \country{USA}}

\aff[e]{Technical University of Denmark, Department of Physics, \city{Kgs. Lyngby}, \country{Denmark}}
 
\date{\today}

\maketitle


\begin{abstract}
We present a method to determine the strain tensor and local lattice rotation with Dark Field X-ray Microscopy. 
Using a set of at least 3 non-coplanar, symmetry-equivalent Bragg reflections, the illuminated volume of the sample can be kept constant for all reflections, facilitating easy registration of the measured lattice variations. 
This requires an oblique diffraction geometry, i.e.~the diffraction plane is neither horizontal nor vertical. 
We derive a closed, analytical expression that allows determination of the strain and lattice rotation from the deviation of experimental observables (e.g.~goniometer angles) from their nominal position for an unstrained lattice.

\end{abstract}


\section{Introduction}

Hierarchically organized crystalline structures are ubiquitous in technological and natural materials such as metals, semiconductors, ceramics, biominerals, geological materials and many others.
The microscopic crystal structure (e.g.~grains or domains) and the atomic-scale defect networks embedded therein determine many of the macroscopic physical and mechanical properties of these materials. 
The need to study crystalline microstructures, and in particular the spatial variations of the strain fields generated by defects, is therefore persistent throughout materials science \cite{Callister2000}.

Diffraction-contrast microscopy techniques have historically been highly effective at mapping strain in crystalline materials. Electron microscopy techniques have mapped strain by Electron Backscatter Diffraction \citeaffixed{Schwartz2009}{EBSD, } and Transmission Electron Microscopy \citeaffixed{Williams2009}{TEM, }.
X-ray techniques are complementary by allowing a larger field of view for \emph{bulk} strain measurements with a resolution that is more coarse spatially, but finer angularly. 
For example, scanning Laue diffraction microscopy can map the 3D strain tensor based on spectrally-resolved XRD of the lattice to measure strain based on distortion of many lattice peaks over long timescales associated with scanning techniques \cite{tamura2003scanning,liu2004three}. Similary, Scanning X-ray Diffraction Microscopy (SXDM) probes the lattice strain by mapping the distortion of single lattice peaks in a spatial raster scan \cite{Zatterin25}.
More recently, Bragg coherent diffraction imaging (BCDI) demonstrated similar advances to solve for the 3D strain tensor field based on the amplitude and phase information from image reconstructions along 3 non-coplanar lattice planes \cite{hofmann2020nanoscale}, capturing high-resolution views of sub-micrometer crystals or grains.

Dark-field x-ray microscopy \citeaffixed{Simons2015}{DFXM, } can be used to map strain fields within bulk crystalline materials over regions of hundreds of micrometers.
As a new technique, however, to date it has been used primarily to measure along only one single Bragg reflection, giving access to up to one longitudinal and two shear strains ---
in order to determine all 6 independent components of the strain tensor, $\tensor{\varepsilon}$, at least 3 non-collinear Bragg reflections must be measured.
In DFXM, measurements often focus on strain fields around isolated lattice defects such as dislocations or coherent small angle (sub)grain boundaries.
In this context, the \emph{local deviation} from the nominal, unstrained crystal lattice is of interest \cite{Poulsen2021}.
Hence the spatial variation of the strain tensor, $\tensor{\varepsilon}$, and the local lattice rotation, $\tensor{w}$, relative to a reference has to be measured\footnote{In order to avoid confusion with the goniometer rotation $\omega$ defined below, we use $\tensor{w}$ to denote the local lattice rotation, whereas the literature often uses $\tensor{\omega}$, e.g.~\citeasnoun{Poulsen2021}.}

Monochromatic diffraction techniques require the sample to be rotated when changing reflections.
Specifically, in a typical DFXM measurement the diffraction plane is fixed to the $x-z$ plane, with line beam illumination perpendicular to this plane. 
Aligning multiple non-coplanar Bragg reflections then requires the sample to be rotated about several axes \citeaffixed{Poulsen2017}{e.g.~$\mu$, $\chi$ and $\phi$, }, which changes the virtual section illuminated by the line beam. 
Registration of the different measurements into a 3D model of the lattice strain in the sample is therefore a challenge.

\citeasnoun{Chung1999} and \citeasnoun{Abboud2017} circumvented this challenge by using white beam Laue diffraction, where several Bragg reflections can be observed without moving the sample.
This method allows for precise measurement of the deviatoric strain, but relies on the relatively poor energy resolution of the detector for the isotropic strain (change of unit cell volume). 

An alternative approach is, for each reflection, to reconstruct the change in Bragg position within the entire gauge volume, and to then register the volumes measured at different reflections. This technique was used by \citeasnoun{Chatelier2024}, who used Bragg Coherent Diffraction Imaging (BCDI) to measure the 3D displacement field within an isolated 350\un{nm} particle of \chem{Ni_{3}Fe}.
A further alternative approach for macroscopic sample sizes could be the ``topo-tomo'' technique \cite{Ludwig2001}. Here the sample is rotated about a selected reciprocal space vector of the unstrained sample, and tomographic methods are used to reconstruct a 3D model of the strain components probed by this reflection. Extending this method to 3 or more non-collinear Bragg reflections, however, is mechanically challenging, as for each reflection a full $360^\circ$ rotation about the corresponding reciprocal lattice vector $\vec{g}$ is needed.

Here we describe a procedure for the observation several non-collinear Bragg reflections for the determination of the complete strain tensor in DFXM and other x-ray diffraction techniques that measure one reflection at a time.
The main idea of this procedure is to facilitate the registration of the volumes probed at the different reflections by ensuring that the same sample volume is illuminated for all reflections. 
In the geometry proposed here, we achieve constant illumination by rotating the sample about an axis perpendicular to the incident beam. 
This is particularly convenient for the case of line beam illumination, as the 3D registration problem is reduced to a 2D problem within the illuminated section, with a known rotation angle between the different measurements.
We choose this rotation axis to be along the laboratory $\hat{z}$-axis (``up'', see Fig.~\ref{fig:z-axis}), and align a reference direction $\vec{h}_{\mathrm{sym}}=(h_{\mathrm{sym}}, k_{\mathrm{sym}}, \ell_{\mathrm{sym}})^T$ of the sample parallel to the rotation axis. 
Observing this reference reflection is not necessary for the procedure described below, as it only serves as a symmetry reference.
Observation does, however, help with alignment of the symmetry axis to the $\omega$ rotation axis.
While the formalism was developed with DFXM in mind, the method can also be used for other experiments performed in the $z$-axis geometry, such as classical section topography \cite{Tran2021,Yoneyama2023} or scanning x-ray diffraction
\cite{Corley-Wiciak2024}.

The paper is organized into two main parts. 
First, Sections \ref{section:general}--\ref{section:determine} discuss the general formalism for the reconstruction of the full strain tensor and lattice rotation from a series of Bragg reflections by classical X-ray diffraction (XRD). 
Our treatment in this section is general and does not rely on a specific diffraction geometry, symmetry or other details of the experiment.
We assume that the deviations of each reflection from their known unstrained reference positions are small and we thus treat those deviations as perturbations upon the main diffracting beam.
In the second part, Sections \ref{section:specific}--\ref{section:examples} discuss how our general approach can be applied to a DFXM experiment using the geometry outlined above, i.e.~by rotating the sample about a single axis which is perpendicular to the incident beam.
This paper focuses on the diffraction geometry and reciprocal-space characteristics of the experiment, leaving the imaging details for a future work \cite{AxelUnpublished,BrinthanUnpublished}.


\section{General formalism}
\label{section:general}

Before examining the specific geometry we use in our experiments, we first share some general observations about measuring the strain tensor with X-ray diffraction, specifically how the strain and lattice rotations can be derived from angular shifts of the peak positions without explicit reconstruction of the distorted unit cell as proposed by \citeasnoun{Schlenker1978}.

In this section, we assume a generalized diffractometer where the reciprocal space vector in the diffraction condition is given by a set of angles $\vec{\xi} = (\xi_1, \xi_2, \ldots \xi_m)$, i.e.~$\vec{g} = \vec{g}(\vec{\xi})$. 
Possible examples include the classical 4-circle diffractometer \cite{Busing1967} where $\vec{\xi} = (2\theta,\omega,\chi,\phi)$, the DFXM diffractometer \cite{Poulsen2017} where $\vec{\xi} = (2\theta, \eta, \mu, \omega, \chi, \phi)$, and many others \cite{Bloch1985,Lohmeier1993,You1999,Bunk2004}.
The $\xi_m$ can be generalized to other observables, e.g.~the photon energy.


\subsection{Defining the Sample \& Goniometer}
\label{sec:21}

In this work, we describe the sample using the ``$\tensor{U}\tensor{B}$''-matrix, as introduced in \citeasnoun{Busing1967} and \citeasnoun{Poulsen2017}. 
In this formalism, the $3\times 3$ matrix $\tensor{B} = (\vec{a}^*, \vec{b}^*, \vec{c}^*)$ contains the basis vectors $\vec{a}^*$, $\vec{b}^*$ and $\vec{c}^*$ of the reciprocal lattice. 
Here we use the convention of \citeasnoun{Schlenker1978} and \citeasnoun{Poulsen2017} where $\tensor{B}$ is a lower triangular matrix.
The orthogonal $3\times 3$ matrix $\tensor{U}$ describes how the sample is oriented relative to the mounting point of the diffractometer \cite{Busing1967,Poulsen2017,Poulsen2021}. 

In the coordinate system attached to the sample mounting point of the diffractometer, the reciprocal space vector $\vec{g}_\diff$ corresponding to the Miller indices $\vec{h} = (h, k, \ell)^T$ is given by  \cite{Poulsen2017}

\begin{align}
    \vec{g}_\diff &= \tensor{U}\tensor{B}  \vec{h}.
    \label{eq:ub}
\end{align}

We refer to this as the \emph{sample coordinate system} \cite{Poulsen2021} and decorate vectors in this coordinate system with the subscript $\diff$. 

For a given reciprocal-space vector to be studied, the diffractometer angles must be set to specific positions $\vec{\xi}$. These positions are not unique, due to the redundancy of the goniometer angles, see e.g.~\citeasnoun{Busing1967}.
Taking the reverse view, given the angle settings of the diffractometer, $\vec{\xi}$, we can calculate the reciprocal space vector, $\vec{g}(\vec{\xi})$, under study \cite{Poulsen2017}.

In this section, we focus on the general aspects of how strain induces shifts in the goniometer angles at which a given reflection is observed, and how this effect can be used to determine the strain tensor.


\subsection{Strain}

For small deformations, the reciprocal lattice in the strained state can be described by 
\begin{align}
    \tensor{B}
    &= 
    \tensor{F}^{-T}  \tensor{B}_0
    \approx
    (\tensor{1} - \tensor{\varepsilon} + \tensor{w}) 
    \tensor{B}_0,
    \label{eq:tensorF}
\end{align}
where $\tensor{F} \approx \tensor{1} + \tensor{\varepsilon} + \tensor{w}$ is a $3\times 3$ tensor and $\tensor{F}^{-T} = (\tensor{F}^{-1})^T$, see e.g.~\citeasnoun{Bernier2011} and \citeasnoun{Poulsen2021}.

Here, the strain tensor $\tensor{\varepsilon}$ is symmetric with 6 free parameters, 

\begin{align}
    \tensor{\varepsilon}
    &=
    \frac{1}{2}
    (\tensor{F}+\tensor{F}^T) - \tensor{1}
    =
    \begin{pmatrix}
        \varepsilon_{xx} & \varepsilon_{xy} & \varepsilon_{xz} \\
        \varepsilon_{xy} & \varepsilon_{yy} & \varepsilon_{yz} \\
        \varepsilon_{xz} & \varepsilon_{yz} & \varepsilon_{zz} 
    \end{pmatrix}
    \label{tensor-epsilon}
\end{align}
because $\varepsilon_{xy}=\varepsilon_{yx}$, etc.

In addition, there may be lattice rotations, described by the antisymmetric matrix $\tensor{w}$ with 3 free parameters,
\begin{align}
    \tensor{w}
    &=
    \frac{1}{2}
    (\tensor{F}-\tensor{F}^T)
    =
    \begin{pmatrix}
        0 & w_z & -w_y \\
        -w_z & 0 & w_x \\
        w_y & -w_x & 0
    \end{pmatrix}.
    \label{tensor-omega}
\end{align}

Different Bragg reflections $\vec{h}=(h,k,\ell)^T$ are sensitive to different parts of the strain and rotation tensors, leading to a (small) deviation of the reciprocal space vector relative to its nominal (unstrained) value, $\vec{g} = \vec{g}_0 + \Delta\vec{g}$, where $\vec{g}_0 = \tensor{U}\tensor{B}_0  \vec{h}$ in the unstrained case, and $\vec{g} = \tensor{U}\tensor{B}  \vec{h}$ in the strained case.

The \emph{shift in reciprocal space} due to strain and lattice rotation is then given by
\cite{Bernier2011,Poulsen2021}
\begin{align}
    \Delta\vec{g}_\diff 
    &= 
    \tensor{U}
    (\tensor{F}^{-T} - \tensor{1})
    \tensor{B}_0
    \vec{h}.
%
%
    \label{eq:Delta-UB}
\end{align}


In an experiment, the reciprocal-space vector will have a small shift in angular peak positions, $\Delta\vec{\xi} = (\Delta\xi_1, \Delta\xi_2, \ldots, \Delta\xi_m)$ compared to the nominal, unstrained peak positions $\vec{\xi} = (\xi_1, \xi_2, \ldots, \xi_m)$,
\begin{align}
    \Delta\vec{g}
    = &
    \vec{g}(\vec{\xi} + \Delta\vec{\xi})
    -
    \vec{g}(\vec{\xi}).
\end{align}

To a linear approximation 
\begin{align}
    \Delta\vec{g}
    & \approx
    \Delta\vec{\xi}
    \nabla\vec{g}
    \\
    &=
    \sum_{p=1}^{m}
    \Delta\xi_p
    \frac{
        \partial \vec{g}
    }{
        \partial \xi_p
    },
    \label{eq:Delta-g-derivative}
\end{align}
where $p$ enumerates the diffractometer angles and $\nabla\vec{g}_{i,j} = \frac{\partial \vec{g}_i}{\partial \xi_j}$ is the $3 \times m$ matrix of gradients of the observed position in reciprocal space with respect to the goniometer angles.


\section{Determining strain and lattice rotation from observed peak shifts}
\label{section:determine}

Eqs.~\ref{eq:Delta-UB} and \ref{eq:Delta-g-derivative} relate the shift of one reciprocal space vector as function of the corresponding goniometer angles. As discussed above, at least 3 non-coplanar Bragg reflections have to be observed in order to calculate the complete strain tensor and lattice rotation.
We therefore generalize these equations by grouping several vectors into (rectangular) matrices.
\begin{align}
    \tensor{H}
    &=
    ( \vec{h}_1, \vec{h}_2, \ldots, \vec{h}_n )
    & & \in \mathbb{R}^3 \times \mathbb{R}^n
    \label{eq:tensorH}
    \\
    \Delta\tensor{\Xi}
    &=
    ( \Delta\vec{\xi}_1, \Delta\vec{\xi}_2, \ldots, \Delta\vec{\xi}_n)
    & & \in \mathbb{R}^3 \times \mathbb{R}^n
    \\
    \Delta\tensor{G}
    &=
    ( \Delta\vec{g}_1, \Delta\vec{g}_2, \ldots, \Delta\vec{g}_n)
    & & \in \mathbb{R}^3 \times \mathbb{R}^n
    \\
    \nabla\tensor{G}
    &=
    ( \nabla\vec{g}_1, \nabla\vec{g}_2, \ldots, \nabla\vec{g}_n),
    & & \in \mathbb{R}^3 \times \mathbb{R}^m \times \mathbb{R}^n
\end{align}
where $n$ is the number of Bragg reflections under study\footnote{Note that this definition of $\tensor{H}$ is different from that of \cite{Poulsen2021}.}.

Then eqs.~\ref{eq:Delta-UB} and \ref{eq:Delta-g-derivative} are generalized to
\begin{align}
    \Delta\tensor{G}_\diff 
    &=
    \tensor{U}
    (\tensor{F}^{-T} - \tensor{1})
    \tensor{B}_0  \tensor{H} 
    \\
    &=
    \Delta\tensor{\Xi}
    \nabla \tensor{G}_\diff.
    \label{eq:allstrain}
\end{align}

For $n>3$ non-coplanar vectors the system is over\-determined and, under the assumption of Gaussian isotropic noise in $\Delta\tensor{G}$, the error in Eq.~\ref{eq:allstrain} is minimized by \cite{Anton2020}
\begin{align}
    \tensor{F}^{-T} - \tensor{1}
    &=
    \tensor{U}^T
    \left[ 
        \Delta\tensor{G}_\diff \,
        (\tensor{B}_0 \tensor{H})^T
    \right]
    \left[ \tensor{B}_0  \tensor{H}  (\tensor{B}_0  \tensor{H})^T \right]^{-1} 
    \\
    &=
    \tensor{U}^{T}
     \left[ (
        \Delta\tensor{\Xi} \,
        \nabla\tensor{G}_\diff
        )  (\tensor{B}_0  \tensor{H})^T \right]
     \left[ \tensor{B}_0  \tensor{H}  (\tensor{B}_0  \tensor{H})^T \right]^{-1},
    \label{eq:final}
\end{align}
where the terms in square brackets are $3\times 3$ matrices.
The strain and rotation components are isolated by being symmetric and antisymmetric, respectively (Eqs.~\ref{tensor-epsilon} and \ref{tensor-omega}).

Eq.~\ref{eq:final} provides a closed, non-iterative formula that connects the angular shifts observed in the diffraction experiment, collected
in the matrix $\Delta\tensor{\Xi}$, to the lattice deformations $\tensor{F}^{-T}-\tensor{1}$ and thus material strains $\tensor{\varepsilon}$. 
It is therefore our central result in this work. 

It can also provide guidance for planning an experiment, specifically for selecting a set of Bragg reflections.
Eq.~\ref{eq:final} will only yield a valid result if the $3\times 3$ matrix $\left[ \tensor{B}_0 \tensor{H} (\tensor{B}_0 \tensor{H})^T\right]$ can be inverted, i.e.~when its determinant is non-zero. 
In the case of 3 reflections, this implies that the determinant of $\tensor{H}$ does not vanish, i.e.~that the the chosen reciprocal space vectors are not collinear.
Furthermore, for numerical stability it is desirable that this determinant is large, in other words that the angle between the reflections is sufficiently large.
Measuring more than 3 reflections will reduce statistical errors through averaging.

To aid with an eventual implementation in a programming language such as python, we also provide the explicit summations of the various tensor products above: The index $i = x,y,z$ labels coordinate axes, the index $j = 1, 2, \ldots, n$ labels reciprocal space vectors or Bragg reflections, and the index $p = 1, 2, \ldots, m$ labels diffractometer angles. Eqs.~\ref{eq:tensorH} to \ref{eq:allstrain} then take the form
\begin{align}
    \tensor{H}_{i,j}
    &=
    h_{i,j}
    \\
    \Delta\tensor{\Xi}_{p,j}
    &=
    \Delta\xi_{p,j}
    \\
    \Delta\tensor{G}_{i,j}
    &=
    \Delta g_{i,j}
    \\
    \nabla\tensor{G}_{i,j,p}
    &=
    \frac{
        \partial g_{i,j}
    }{
        \partial \xi_p
    }
    \\
    (\Delta\tensor{\Xi}\,  \nabla\tensor{G})_{ij}
    &=
    \sum_{p=1}^{m}\Delta\xi_{p,j}
    \frac{
        \partial g_{i,p}
    }{
        \xi_j
    }.
\end{align}
The product between $(\Delta\tensor{\Xi}\, \nabla\tensor{G})$ and $(\tensor{B}_0 \tensor{H})^T$ is summed over the index $j$ which labels the reciprocal space vectors. 


\section{Specific geometry for Dark Field X-ray Microscopy}
\label{section:specific}

We now turn to discussing the specific experimental geometry used in our experiments.
For DFXM, recording the goniometer and associated strain information in 2D images raises the problem of reconstructing a 3D sample volume from 2D projections. Since the crystal orientation differs significantly between lattice peaks, this implies that the voxels of sample corresponding to each pixel differ significantly, requiring image registration to consolidate  the volumes obtained from different Bragg reflections into each specific sample region over all orientations $\vec{\xi}_i$.
 
In this section, we address the image registration challenge by proposing a measurement geometry that can ensure the same sample volume is illuminated for all non-coplanar orientations.
Constant illumination is achieved by reducing the sample movement to a single rotation $\omega$ about an axis perpendicular to the incident beam. 
This diffraction geometry is a 1S+2D geometry \citeaffixed{Bloch1985,Bunk2004}{1 sample rotation, 2 detector rotations, see }, even though two additional sample rotations are needed to align the chosen symmetry axis parallel to the main axis of rotation ($\omega$ axis). 

The configuration is similar to the $z$-axis geometry \cite{Bloch1985,Bunk2004}, the 3DXRD geometry \cite{Poulsen2001,Jakobsen2006} and diffraction contrast tomography \citeaffixed{King2008,Ludwig2009}{DCT, }.
As these additional axes are not moved during this experiment we do not consider them explicitly and assume that they are absorbed into $\tensor{U}$. 
Full geometry calculations can be found in the literature, e.g.~\citeasnoun{Poulsen2017}.
In this geometry, there are two detector angles, $2\theta$ and $\eta$, and one sample angle $\omega$, such that $\vec{\xi}=(\omega,\eta,2\theta)$, see Fig.~\ref{fig:z-axis}.


Contrary to most DFXM experiments, in our proposed geometry the diffracted beam is not confined to the vertical plane. 
Instead it is rotated out of this plane by an angle $\eta$ \cite{Poulsen2017}.
A similar approach relying instead on a 2S+1D geometry (2 sample and 1 detector rotation) was used in scanning X-ray microscopy experiments where a nano-beam was raster-scanned across the sample \cite{Richter2022,Corley-Wiciak2023b,Corley-Wiciak2024}.

A unique advantage of this geometry is that the selected symmetry-equivalent reflections can all be observed at the same detector position, and that only a single sample rotation is required to switch between them.
As we show below, several such detector positions exist.
The number of measurements can thus be increased by collecting the same reflections at several possible detector positions.
Friedel pairs $(h,k,\ell)$ and $-h,-k,-\ell)$ can be collected to further improve the statistics.

DFXM experiments are often carried out with a line-focused beam, allowing access to the full 6D dataset for mechanical deformation in the sample (3D reciprocal + 3D real space).
The method proposed in this work allows for registration even within the line-focused illuminating beam because the measured 3D volume is all limited to the same 3D plane that rotates about a single axis that is orthogonal to the incident beam, $\omega$ (see Fig.~\ref{fig:z-axis}).
We note that there are downsides to this geometry. 
Notably, only certain reflections with $\alpha > \theta$ can fulfill the diffraction condition, where $\alpha$ is the angle between the reflection under study and the $\omega$ axis of rotation (see Eq.~\ref{eq:simplified-diffraction-condition}).

\subsection{Summary of the different coordinate systems}

Following \citeasnoun{Poulsen2017}, we define a sequence of coordinate system as follows:

\begin{itemize}
\item The \emph{laboratory coordinate system}, indicated by the subscript $\vec{g}_\lab$ \citeaffixed{Poulsen2017}{see Fig.~\ref{fig:z-axis} and } is fixed and does not move with any goniometer rotations. The direction of the incident beam is along the positive $\hat{x}_\lab$ axis. The $\hat{y}_\lab$ axis is horizontal to port (left when seen along the beam axis). The $\hat{z}_\ell$ axis is up.

\item The \emph{sample coordinate system} as defined by \citeasnoun{Poulsen2021}, is attached to the gonimeter's innermost sample rotation (here the $\omega$ axis), indicated by the subscript $\vec{g}_\diff$.
The sample's orientation, defined by the matrix $\tensor{U}$ is fixed in this coordinate system.
For this experiment, only the goniometer rotation $\omega$ about the $\hat{z}_\lab$ axis is used. 
All other rotation angles are set to zero. 
Thus the matrix $\tensor{\Gamma}$ as used by \citeasnoun{Poulsen2021} is given by $\tensor{\Gamma} = \tensor{R}_z(\omega)$.

\item Miller indices in the \emph{crystal coordinate system}. Here we use the symbol $\vec{h} = \begin{pmatrix}h,k,\ell\end{pmatrix}^T$ without subscript.
\end{itemize}

These are related by
\begin{align}
    \vec{g}_\diff 
    &= \tensor{U}\tensor{B} \vec{h}
    \\
    \vec{g}_\ell &= \tensor{\Gamma} \tensor{g}_\diff
     =\tensor{R}_z(\omega)  \vec{g}_\diff,
    \label{eq:diff-to-lab}
\end{align}

\subsection{Defining the Incident and Diffracted Beams}

In the laboratory coordinate system, the incident beam travels along the positive $\hat{x}_\lab$ axis,

\begin{equation}
  \vec{k}_{\mathrm{in},\lab} = \frac{2\pi}{\lambda} \hat{x}_\lab,
\end{equation}
where $\lambda$ is the X-ray wavelength.

\begin{figure}
	\begin{center}
		\resizebox{\columnwidth}{!}{
			\includegraphics[angle=0]{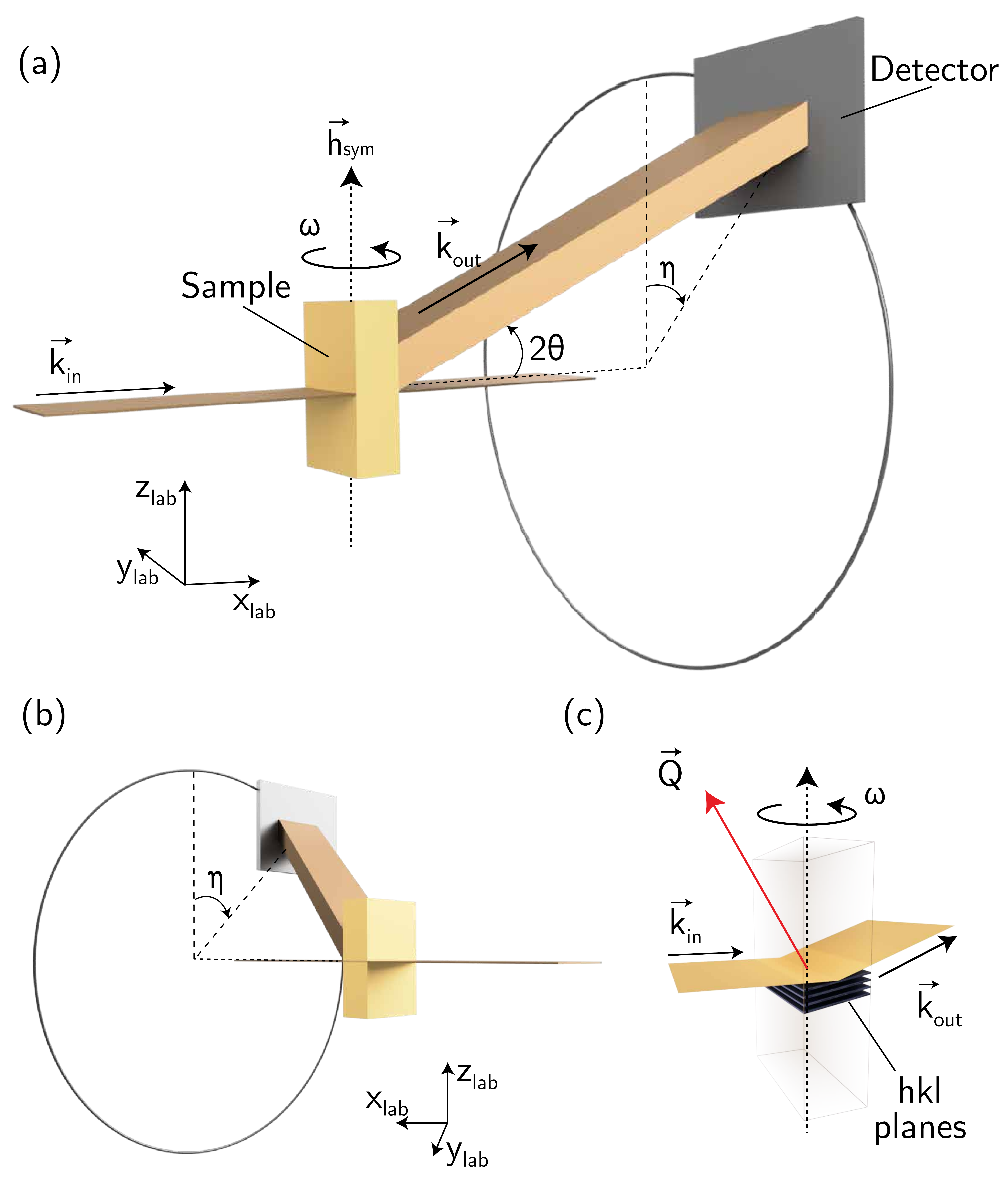}
		}
	\end{center}
	\caption{Diffraction geometry: The incident beam $\vec{k}_{\mathrm{in}}$ travels along the $+\hat{x}$ axis. The diffracted beam is described by the angles $2\theta$ with the $\hat{x}$ axis and $\eta$ with respect to the $xz$ plane. 
	}	
	\label{fig:z-axis}
\end{figure}

Rather than using the angles $\delta$ and $\gamma$ to describe the detector rotations \cite{Bloch1985,Lohmeier1993,You1999,Bunk2004}, we employ the description used for large area detectors, for example in 3DXRD \cite{Poulsen2001}, also used for DFXM \cite{Poulsen2017}. Here the beam is first rotated by $2\theta$ about the $-\hat{y}_\lab$ axis, then by $\eta$ about the incident beam axis, see Fig.~\ref{fig:z-axis}.
The diffracted beam thus is given by
\begin{align}
  \vec{k}_{\mathrm{out},\lab}
  &= \tensor{R}_x(\eta)  \tensor{R}_y(-2\theta)  \vec{k}_{\mathrm{in},\lab}
  \\
  &= k \begin{pmatrix}
    \cos(2\theta) \\
    -\sin(2\theta) \sin(\eta) \\
    \sin(2\theta) \cos(\eta).
    \end{pmatrix}_\lab
\end{align}

Here $\tensor{R}_{x,y,z}(\alpha)$ are (right-handed) rotations about the $x,y,z$ axes through the angle $\alpha$, respectively \cite{Poulsen2017}. The scattering vector (in the laboratory coordinate system) is given by
\begin{align}
  \vec{Q}_\lab
  &= \vec{k}_{\mathrm{out},\lab} - \vec{k}_{\mathrm{in},\lab}
  \\
  &= 2k \sin(\theta) \begin{pmatrix}
    -\sin(\theta) \\
    -\cos(\theta) \sin(\eta) \\
    \cos(\theta) \cos(\eta) 
  \end{pmatrix}_\lab.
  \label{eq:q}
\end{align}

In particular
\begin{align}
  \left| \vec{Q}_\lab \right|
  &= 2k \sin(\theta).
  \label{eq:qlen}
\end{align}

\subsection{Simplified Sample Goniometer}

To obtain the scattering vector $\vec{Q}_\diff$ in the sample coordinate system, this vector is rotated about the $\hat{z}$ axis by $-\omega$,
(see Fig.~\ref{fig:z-axis} and Eq.~\ref{eq:diff-to-lab}).

\begin{align}
    \vec{Q}_\diff(&\omega, \eta, \theta)
    =
    \tensor{R}_z(-\omega)  \vec{Q}_\lab
    \\
    &=
    2 k \sin(\theta) 
    \begin{pmatrix}
    -\cos(\omega) \sin(\theta) - \sin(\omega) \cos(\theta) \sin(\eta) \\
     \sin(\omega) \sin(\theta) - \cos(\omega) \cos(\theta) \sin(\eta) \\
    \cos(\theta) \cos(\eta)
    \end{pmatrix}_\diff.
\end{align}

Bragg's law is fulfilled when this scattering vector is equal to the reciprocal lattice vector of an allowed Bragg reflection,
\begin{align}
    \vec{Q}_\diff
    &=
    \vec{g}_\diff =
    \tensor{U}\tensor{B}\vec{h}.
\end{align}

\subsection{Choice of reciprocal lattice vectors}

We are interested in a series of reciprocal space vectors of the \emph{undistorted crystal} that are related by symmetry, specifically vectors that can be transformed into each other by rotation about a \emph{symmetry axis} $\vec{h}_\sym =\begin{pmatrix} h_\sym, k_\sym, \ell_\sym \end{pmatrix}^T$. 
The corresponding reciprocal space vector in the sample coordinate system is $\vec{g}_{\sym,\diff} = \tensor{U}\tensor{B}_0 \vec{h}_\sym$.
This axis is used as a reference only.
In general we do not observe diffraction from this reciprocal space vector.

We assume the undistorted sample to be mounted such that this symmetry axis is parallel to the goniometer $\omega$ axis, i.e. the matrix $\tensor{U}$ is chosen such that
\begin{align}
  \tensor{U}  \tensor{B}_0  \vec{h}_\sym
  & = \begin{pmatrix} 0 \\ 0 \\ g_\sym \end{pmatrix} = \vec{g}_{\sym,\diff}.
  \label{eq:gsym}
\end{align}
As the $\omega$ axis is parallel to $\hat{z}_\diff$, the $\hat{z}$ axes of the laboratory and sample coordinate systems are identical, $\hat{z}_\diff =\hat{z}_\lab$ axis for all settings of $\omega$.
Consequently, $\omega$ rotations do not change $\vec{g}_{\sym,\lab}$ and $\vec{g}_{\sym,\diff}=\vec{g}_{\sym,\lab}$.

As the $\omega$ axis is perpendicular to the incident beam direction, the illuminated volume of the sample remains unchanged when the sample is rotated about the $\omega$ axis --- especially when the incident beam is focused to a line for section topography \cite{Tran2021,Yoneyama2023}.

Next, we select a family of symmetry-equivalent reflections $\vec{h}_n = (h_n, k_n, \ell_n)^T$ that can be transformed into each other via rotations about the symmetry axis or reflection by mirror planes parallel to the symmetry axis. 
At least 3 symmetry-equivalent reflections are required. 
Therefore the lattice system has to be orthorhombic, tetragonal, rhombohedral, hexagonal, or cubic.
Examples are listed in Table~\ref{tab:reflections}.

\begin{table}
\caption{Examples of groups of symmetry-equivalent reflections $(h,k,\ell)^T$ that can be transformed into each other via rotations about a symmetry axis. C denotes cubic crystal symmetry, T tetragonal, O orthorhombic and H hexagonal.}
\begin{tabular}{lccc}
Crystal & Symmetry axis     & Reflections \\
symmetry &  $(h_\sym, k_\sym, \ell_\sym)$ & $(h_n, k_n, \ell_n)$ \\
\hline
C, T                    & $(h_\sym,h_\sym,0)$      & $(2,0,\pm 2)$, $(0, 2, \pm2)$ \\
                          &                           & $(3,1,\pm 1)$, $(1, 3, \pm1)$ \\ 
C                        & $(h_\sym,h_\sym,h_\sym)$  & $(2,-2,0)$, $(0, 2, -2)$, $(-2, 0, 2)$\footnote{These reflections are co-planar and do not allow for the determination of all coefficients of the strain tensor.} \\
                          &                           & $(4, 0, 0)$, $(0, 4, 0)$, $(0, 0, 4)$ \\
                          &                           & $(3,1,1)$, $(1,3,1)$, $(1,1,3)$ \\
C, T, O                & $(0,0,\ell_\sym)$         & $(\pm 1, \pm 1, 0)$ & \\
C, T                    & $(0,0,\ell_\sym)$         & $(\pm 3, \pm 1, 1)$, $(\pm 1, \pm 3, 1)$\\                                              
H                        & $(0,0,\ell_\sym)$         & $(\pm h,0,\ell)$, $(0,\pm h,0,\ell)$, $(\pm h, \mp h, \ell)$ \\
                        &                            & $(\pm h, \pm h, \ell)$, $(\pm h, \mp 2h, \ell)$, $(\pm 2h, \mp h, \ell)$
\end{tabular}
\label{tab:reflections}
\end{table}

For simplicity, we consider only one of these vectors, $\vec{g}$, which we rotate about the $\hat{z}_\lab$ axis such that it lies within the $y_\lab$-$z_\lab$ plane.
Let
\begin{align}
  \tensor{UB}_0 \begin{pmatrix} h \\ k \\ \ell \end{pmatrix}
  &=
  \begin{pmatrix} g_{x,\diff} \\ g_{y,\diff} \\ g_{z,\diff} \end{pmatrix} = \vec{g}_\diff
  \\
  &=
  \tensor{R}_z(-\omega_0) \begin{pmatrix}0 \\ g_\perp \\ g_{z,\diff} \end{pmatrix},
\end{align}
with $g_\perp \sin(\omega_0) = g_{x,\diff}$ and $g_\perp \cos(\omega_0) = g_{y,\diff}$.

The different symmetry-equivalent reciprocal space vectors then differ only by their respective values of $\omega_0$. Therefore the following calculations are carried out only for one representative vector $\vec{g}_\diff$, taking into account its orientation $\omega_0$.


\subsection{Diffraction condition}

The $\omega$-axis of the goniometer then rotates this vector about the $\hat{z}_\lab$ axis, such that the reciprocal space vector in the laboratory coordinate system is given by
\begin{align}
  \vec{g}_\lab
  &=
  \tensor{R}_z(\omega-\omega_0)  \begin{pmatrix}0 \\ g_\perp \\ g_z \end{pmatrix}
  \label{eq:g-lab}
  \\
  & =
  \begin{pmatrix}
    -\sin(\omega-\omega_0) g_\perp \\
    \cos(\omega-\omega_0) g_\perp \\
    g_z
  \end{pmatrix}
  \label{eq:g}
\end{align}

The task is now to find angles $\theta$, $\eta$ and $\omega$ such that Eq.~\ref{eq:g} equals Eq.~\ref{eq:q}.
$\theta$ is found by comparing the length of the two vectors\footnote{The generalized, but equivalent, equation system was solved by \citeasnoun{xrdsimulator} for an arbitrary orientation of the $\omega$ axis. Here, we provide an alternative derivation, valid for the special case when the $\omega$ axis is kept parallel to $\vec{z}_l$.}, eq.~\ref{eq:qlen}.

\begin{align}
  2k \sin(\theta)
  &= \left| \vec{g} \right|
  = \sqrt{g_\perp^2 + g_z^2}.
  \label{eq:solve-theta}
\end{align}
This is Bragg's law, as $d = 2\pi/\left|\vec{g}\right|$.

Left-multiplying equations~\ref{eq:g-lab} and~\ref{eq:q} with $\tensor{R}_z(-\omega+\omega_0)/\left| \vec{g} \right|$ yields

\begin{equation}
\begin{split}
   &\frac{1}{ \left|\vec{g}\right|}
   \begin{pmatrix} 0 \\ g_\perp \\ g_z \end{pmatrix}
 =
   \begin{pmatrix} 0 \\ \sin(\alpha) \\ \cos(\alpha) \end{pmatrix} \\
 & =
   \tensor{R}_z(-\omega+\omega_0) \begin{pmatrix} -\sin(\theta) \\ -\cos(\theta)\sin(\eta) \\ \cos(\theta) \cos(\eta) \end{pmatrix}
\\
   & =
   \begin{pmatrix}
   -\cos(\omega-\omega_0) \sin(\theta) - \sin(\omega-\omega_0) \cos(\theta) \sin(\eta) \\
   -\cos(\omega-\omega_0) \cos(\theta) \sin(\eta) + \sin(\omega-\omega_0) \sin(\theta) \\
   \cos(\theta) \cos(\eta)
   \end{pmatrix}
   \label{eq:simplified-diffraction-condition}
\end{split}
\end{equation}

For simplicity we assume $0 < \alpha < \pi/2$.
An additional set of solutions can be found for $-\pi/2 < \alpha < 0$. This provides an opportunity for additional measurements.
Solving this yields 

\begin{align}
	\omega - \omega_0
	& =
	\mp \arctan\left(
		\frac{
	 		\sin(\theta)
	 	}{
			\sqrt{\cos^2(\theta) - \cos^2(\alpha)}
		}
	\right)
	\\
	\eta
	& =
	\pm
	\arctan\left(
		\frac{
			\sqrt{\cos^2(\theta) - \cos^2(\alpha)}
		}{
			\cos(\alpha)
		}
	\right).
\end{align}

Note that due to the symmetry requirement, all reflections share the same values of $\theta$ and $\alpha$, and therefore $\omega-\omega_0$ and $\eta$. Only the value of $\omega_0$ (and therefore $\omega$) varies from reflection to reflection.
In other words, all reflections can be measured without moving the detector (except scanning around the nominal position). 

Valid solutions are only found for $\cos(\theta) > \cos(\alpha)$, i.e. $\alpha > \theta$. We use the $\mathrm{arctan2}$ function, which correctly determines the sector and avoids division by zero.

\begin{align}
	\omega - \omega_0
	& =
	\mathrm{arctan2}\left(
	 	\sin(\theta),
		\mp \sqrt{\cos^2(\theta) - \cos^2(\alpha)}
	\right)
    \label{eq:solve-omega}
	\\
	\eta
	& =
	\pm
	\mathrm{arctan2}\left(
		\sqrt{\cos^2(\theta) - \cos^2(\alpha)},
		\cos(\alpha)
	\right).
    \label{eq:solve-eta}
\end{align}


\subsection{Linearized shifts}

In our diffraction geometry, Eq.~\ref{eq:Delta-g-derivative} is explicitly
\begin{align}
    \Delta\vec{g}
    = &
    2k \sin(\theta) \Delta\omega 
    \begin{pmatrix}
    \sin(\theta) \sin(\omega) - \cos(\theta) \cos(\omega) \sin(\eta) \\
    \sin(\theta) \cos(\omega) + \cos(\theta) \sin(\omega) \sin(\eta) \\
    0
    \end{pmatrix}
\nonumber \\ &
    -
    k \sin(2\theta)
    \Delta\eta
    \begin{pmatrix}
    \sin(\omega) \cos(\eta)  \\
    \cos(\omega) \cos(\eta)  \\
    \sin(\eta) 
    \end{pmatrix}
\nonumber \\ &
    + 2k
    \Delta\theta
    \begin{pmatrix}
    -\sin(2\theta)\cos(\omega) - \cos(2\theta) \sin(\omega) \sin(\eta) \\
    \sin(2\theta) \sin(\omega) - \cos(2\theta) \cos(\omega) \sin(\eta)  \\
    \cos(2\theta) \cos(\eta)
    \end{pmatrix}.
    \label{eq:taylor}
\end{align}

Note again that $\theta$ and $\eta$ have the same value for all reflections under consideration, only the value of $\omega$ varies.


\section{Examples}
\label{section:examples}

To demonstrate how the method could be put into practice, we present an example specifically chosen for the Dark Field Microscopy instrument on ID03 of the ESRF \cite{Isern2024}.
For the detector geometry at this beamline, we need $-\pi/2 < \eta < 0$ \cite{Poulsen2017,Isern2024}.
A common application of DFXM is the study of dislocation and dislocation structures in metals.
One of the metals most studied by DFXM is Aluminium \cite{Simons2015,Dresselhaus2021,Yildirim2023}.

Aluminium is FCC cubic with lattice constant $a=4.0495\,$\AA.  
We use $\vec{h}_\sym = (2,\bar{2},0)^T$ as reference direction and measure the reflections $\vec{h}_{1\ldots 4} = (2, 0, \pm 2)^T$ and $(0, \bar{2}, \pm 2)^T$.
The angle between the reference direction and the reciprocal lattice vectors of interest is $\alpha= \arccos(1/2) = 60^\circ$.

Let 
\begin{align}
	\tensor{U}
	& =
	\begin{pmatrix}
	\frac{1}{\sqrt{2}} &  \frac{1}{\sqrt{2}} & 0 \\
	0 &  0 & 1 \\ 
	\frac{1}{\sqrt{2}} & -\frac{1}{\sqrt{2}} & 0 \\ 
	\end{pmatrix},
\end{align}
such that the symmetry axis $(2,\bar{2},0)^T$ is parallel to $\hat{z}$,
\begin{align}
	\tensor{UB} \begin{pmatrix}2 \\ -2 \\ 0 \end{pmatrix} 
	& = \frac{2\pi}{a}
	\begin{pmatrix}0 \\ 0 \\ 2\sqrt{2}\end{pmatrix}
\end{align}

The reciprocal lattice vectors in the laboratory coordinate system are then
\begin{align}
	\tensor{UB} \begin{pmatrix}2 \\ 0 \\ 2\end{pmatrix} 
	& = \frac{2\pi}{a}
	\begin{pmatrix} \sqrt{2} \\ 2 \\ \sqrt{2} \end{pmatrix};
	& \omega_0 &= 35.264^\circ
	\\
	\tensor{UB} \begin{pmatrix} 2 \\ 0 \\ -2 \end{pmatrix}
	& = \frac{2\pi}{a}
	\begin{pmatrix} \sqrt{2} \\ -2 \\ \sqrt{2} \end{pmatrix};
	& \omega_0 &= 144.736^\circ
	\\
	\tensor{UB} \begin{pmatrix} 0 \\ -2 \\ 2 \end{pmatrix}
	& = \frac{2\pi}{a}
	\begin{pmatrix} -\sqrt{2} \\ 2 \\ \sqrt{2} \end{pmatrix};
	& \omega_0 &= -35.264^\circ
	\\
	\tensor{UB} \begin{pmatrix} 0 \\ -2 \\ -2 \end{pmatrix}
	& = \frac{2\pi}{a}
	\begin{pmatrix} -\sqrt{2} \\ -2 \\ \sqrt{2} \end{pmatrix};
	& \omega_0 & = -144.736^\circ
\end{align}

At $E=19.1\un{keV}$, $\theta \approx 13.103^\circ$, $2\theta \approx 26.206^\circ$. $\omega-\omega_0 \approx 15.175^\circ$, $\eta \approx -59.112^\circ$.


\section{Conclusion}

In conclusion, we present an experimental method and data analysis process to derive the full strain tensor from a series of Dark Field X-ray Microscopy measurements. 
The method relies on measuring small angular deviations between strained and unstrained parts of the sample, e.g.~around dislocations in near-perfect crystals.
Treating these deviations as perturbations, we derive a non-iterative closed formula for calculating the corresponding lattice strain and rotation relative to the unstrained reference lattice.

By choosing an oblique diffraction plane and rotating the sample about a symmetry axis, several symmetry-equivalent, but non-coplanar, Bragg reflections can be measured without moving the detector.
This greatly facilitates the registration of the gauge volume within the sample, which is necessary for the 3D reconstruction of the strain field throughout the sample volume.
We explicitly calculate the sample and detector angles for this oblique geometry.

Future work will extend our approach to take into account the imaging aspect of DFXM, in particular the fact that strained and unstrained regions of the sample can be measured simultaneously.

While this study focused on a single crystal, integrating DFXM with grain-resolved 3DXRD paves the way for full strain tensor mapping in polycrystalline materials. In such cases, 3DXRD can identify grains of interest and their orientations, enabling targeted DFXM scans \cite{Shukla2025}. Ongoing developments in X-ray optics for DFXM measurements, particularly diamond-based lenses \cite{Seiboth2017,Celestre2022} and higher energy X-rays, will expand the accessible \textit{hkl} range, supporting broader application of this method to studies with complex, industrially relevant polycrystalline materials.



\ack{Acknowledgements}:  

C.Y.~acknowledges the financial support by the ERC Starting Grant ``D-REX'' nr.~101116911.
N.A.H, A.A.W.C., F.T.F., S.B. and H.F.P.~acknowledge support  from ERC Advanced Grant nr.~885022 and from the Danish ESS lighthouse on hard materials in 3D, SOLID.
This study has received financial support from the European Union:
AddMorePower (GA 101091621).


\referencelist[Oblique]

\end{document}